# Revealing the grain structure of graphene grown by chemical vapor deposition

Péter Nemes - Incze[1,3,*], Kwon Jae Yoo[2,3], Levente Tapasztó[1,3], Gergely Dobrik[1,3], János Lábár[1], Zsolt E. Horváth[1,3], Chanyong Hwang[2,3], László Péter Biró[1,3]

*1 Research Institute for Technical Physics and Materials Science, PO Box 49, 1525 Budapest, Hungary*

*2 Center for Advanced Instrumentation, Division of Industrial Metrology, Korea Research Institute of Standards and Science, Yuseong, Daejeon 305-340, Republic of Korea*

*3 Joint Korean-Hungarian Laboratory for Nanosciences (JKHLN), PO Box 49, 1525 Budapest, Hungary*

**Abstract**

The physical processes occurring in the presence of disorder: point defects, grain boundaries, etc. may have detrimental effects on the electronic properties of graphene. Here we present an approach to reveal the grain structure of graphene by the selective oxidation of defects and subsequent atomic force microscopy analysis. This technique offers a quick and easy alternative to different electron microscopy and diffraction methods and may be used to give quick feedback on the quality of graphene samples grown by chemical vapor deposition.

[*] Corresponding author:
tel: +36-1-3922526
fax: +36-1-3922226
email: nemes@mfa.kfki.hu
URL: www.nanotechnology.hu

The ease with which graphene can be prepared by micromechanical cleavage[1], has made this exiting two dimensional carbon crystal available to a wide community of researchers and thus has contributed significantly to the development of this research field. On the same note, the relative ease with which graphene can be grown on metal surfaces by chemical vapor deposition (CVD)[2,3,4] is a major advantage in the large scale adoption of graphene in industry. Some applications which seem particularly close are the ones that use graphene as a transparent electrode material, in solar cells and displays[5,6,7,8]. However, there are significant differences in the physical properties of graphene samples prepared by cleaving graphite (HOPG, Kish graphite or high grade natural graphite) and samples prepared by CVD, attributable to the microstructure of these materials. Grain boundaries and in a more general case, line defects may dramatically influence the electronic[9,10] and mechanical[11,12] behavior of graphene. Important among these, from the point of view of electronic applications, is that the charge carrier mobility of CVD graphene samples can be orders of magnitude lower[2] than theoretical values[13] or the data reported for cleaved graphene[14,15]. The biggest culprit in this may be the physical processes occurring at grain boundaries. Various groups have mapped the grain structure of CVD grown graphene by transmission electron microscopy (TEM)[11,16,17] and have shown that the material is a patchwork of large angle grain boundaries, with the characteristic grain size in the 100 nm range. Improving these properties is a goal best achieved by finding the growth parameters that result in samples with the highest crystallinity and the smallest number of grain boundaries. However, this is usually a prolonged, empiric and iterative process of analyzing the grown graphene sample, tweaking growth parameters and performing the CVD growth. Here we report a method that gives a quick assessment of the grain structure of graphene, which if employed in the CVD optimization procedure, has the potential to substantially speed up this process.

Graphene samples used in this study were prepared by the CVD technique pioneered by the group of Ruoff[2,18]. After the graphene growth experiments, the presence of graphene flakes on the copper surface was demonstrated by Raman spectroscopy[2,19] and the samples were subjected to the grain boundary etching procedure.

It is well documented that when heating graphite in an oxygen containing atmosphere, at temperatures below 875°C, oxidation only occurs at defect sites[20] such as point defects, dislocations and grain boundaries. At these defect sites the carbon gasification reaction has lower activation, energy, determined by the alternating pentagon – heptagon structure of the grain boundary[11,16,18]. Thus, the defects can be revealed by etching away the carbon in their close vicinity. We have exploited this effect to reveal the grain structure of graphene. Three parameters can be chosen to control the rate of carbon removal: temperature, heating time and oxygen concentration. To keep the experimental conditions as simple as possible, we have chosen to perform the etching reaction in air, using a temperature of 500°C. This temperature is far away from the values where gasification of carbon atoms occurs all through the graphite basal plane[20] and can be easily obtained in a resistively heated furnace. This leaves as a free parameter the etching time. However, before attempting the heat treatment, the graphene has to be transferred from the copper to another substrate, because the copper itself would oxidize and complicate the interpretation of experimental results. Si wafers with a $SiO_2$ capping layer is the obvious choice for a target substrate, but it's not suitable for our purposes. The reason being that, as our experimental results and that of Liu et al.[21] have shown, on $SiO_2$ the nucleation of etch sites occurs not only at the location of defects but stochastically throughout the graphene layer. The reason for this may be the reduction in the energy barrier for carbon removal due to the local curvature of graphene on a rough $SiO_2$ support, or the presence of charged impurities in the oxide[21]. Taking these considerations into account we have chosen mica as an alternative substrate, because it is atomically flat and of high crystallinity, so we can avoid a high local curvature[22] and having charged impurities in the vicinity of the graphene. We have chosen the widely used graphene transfer technique using PMMA[23]. A $CuCl_2$ solution was used to etch away the copper[18].

Having fixed the oxidation temperature and the oxygen concentration, we have increased the heat treatment time in 5 minute increments, taking atomic force microscope (AFM) images of the sample after each increment. At 30 minutes of etching, the grain boundaries become visible in the AFM images as trenches, with roughly 20 nm widths, crisscrossing the graphene (see Figure 1). These etch trenches are not to be confused with wrinkles in the graphene[2], which occur during growth or the transfer process (Figure 1a).

To make sure that the etchant does not introduce defects in the graphene, we have performed a control experiment by preparing graphene layers on mica from graphite (HOPG) using exfoliation. This high grade graphite has grain sizes in the 100-1000 µm range and very few point defects and thus should be unaffected by the oxidation treatment. After exfoliation the control sample was also placed in the copper etchant for the duration of a typical etch (~30 min) and rinsed with DI water and was kept at 500$^o$C in air for 30 minutes and measured by AFM. This heat treatment left this sample unaffected (Figure 1c). This way we could test that the copper etchant does not introduce additional defects in the graphene and that the etch trenches observed on CVD grown graphene are indeed intrinsic to the CVD grown sample.

After performing this simple heat treatment and subsequent AFM measurements we obtain a map of the grain boundaries, which can be further processed by image analysis software to yield the areas enclosed by the etch trenches. This can be used to construct a histogram of the characteristic size of grains (Figure 2). The characteristic size in the sample discussed here is a few 100 nm, with large grains of 1 µm or more being very rare. While smaller grains show up with increasing frequency in the sample we see a dip in the histogram for very small grains. This can be explained by the fact that grains which are only tens of nanometers in size are not detectable with this technique, because they get etched away completely. Further information can be extracted by this technique if we measure each grain with atomic resolution using contact mode AFM[24]. By taking the Fourier transform of each atomic resolution image we can deduce the crystallographic orientation of each grain and plot a histogram of the relative angle between adjacent grains (see Figure 2). We have to mention that due to tip convolution effects the grain boundaries themselves cannot be imaged directly in a pristine graphene sample. The size of the grains and the fact that large angle grain boundaries dominate the sample are in agreement with TEM measurements on samples produced by similar CVD experiments[11,16].

The ease with which atomic resolution AFM images can be acquired on graphene under ambient conditions is surprising. Therefore, we propose atomic force microscopy as a practical tool for analyzing the atomic structure of graphene, which could be a versatile alternative to other microscopic approaches [25]. Combining this tool with the etching procedure described above we can obtain information on the two most important parameters of the graphene grain structure: the size distribution and the angles of the grain boundaries. One important aspect of this mapping technique is that the kind of contact mode AFM used here is standard equipment in any nanoscience laboratory, in contrast to a low energy, high resolution TEM required for dark field grain boundary mapping[11,16,17]. As described above, our grain boundary analysis method is easy to implement and can give quick feedback for researchers involved in the CVD growth of graphene, helping to bring closer the goal of growing graphene with high crystallinity and better control over the microstructure.

**Acknowledgements**

This work has been conducted within the framework of the Joint Korean-Hungarian Laboratory for Nanosciences (JKHLN), the Converging Research Center Program through the Ministry of Education, Science and Technology (2010K000980) and the OTKA-NKTH grant no. K67793.

**Figures**

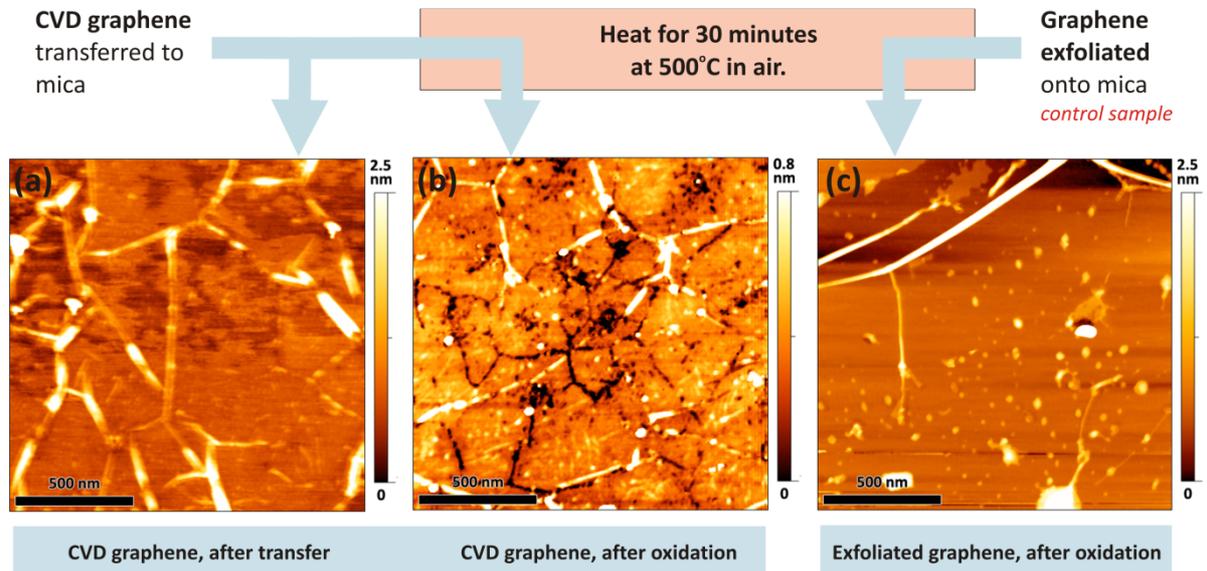

**Figure 1.** The process flow of the sample preparation and tapping mode AFM images of the graphene samples: (a) pristine CVD graphene sample transferred to mica, showing wrinkles; (b) CVD graphene on mica after oxidation, black lines correspond to etch trenches; (c) control sample: graphene exfoliated onto mica from high grade graphite, after exposure to the copper etchant and subsequent oxidation.

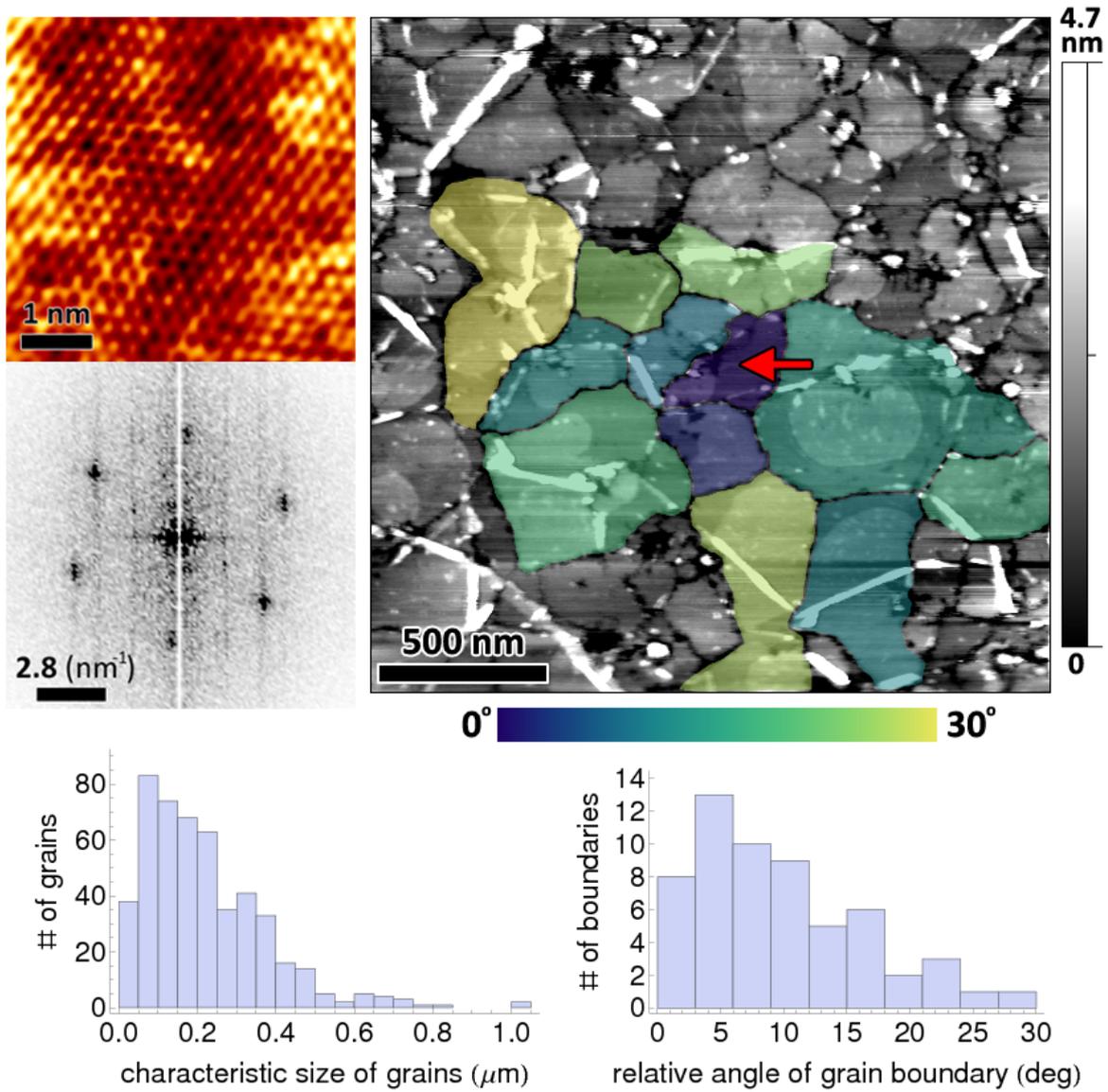

**Figure 2.** AFM image of CVD graphene showing the etch trenches. By comparing the Fourier transform of atomic resolution AFM images of individual grains, one can make a false color map of the crystallographic orientation of the grains relative to a given direction. Two small images to the left show one such atomic resolution image and the Fourier transform thereof. The atomic resolution image was recorded by contact mode AFM on the grain shown by the red arrow. Histograms at the bottom show the distribution of the grain size (left) and the relative angle of the grains forming the boundaries (right). This data is a compilation of multiple AFM mapping measurements.